\documentclass[showpacs,amsmath,amssymb,aps,preprint,nofootinbib]{revtex4-1}

\usepackage{graphicx}
\usepackage{dcolumn}
\usepackage{color}
\usepackage{float}

\usepackage{hyperref}
\hypersetup{%
colorlinks=true,%
linkcolor=blue,
citecolor=blue,
urlcolor=magenta
}
\newcommand{\sumint}{\mbox{$\sum$}\kern-2.7ex\int}

\def\gtsim{\mathrel{\hbox{\raise0.2ex
\hbox{$>$}\kern-0.75em\raise-0.9ex\hbox{$\sim$}}}}
\def\ltsim{\mathrel{\hbox{\raise0.2ex
\hbox{$<$}\kern-0.75em\raise-0.9ex\hbox{$\sim$}}}}

\begin{document}


\title{Gravitational waves from domain wall collapses and dark matter in the SM with a complex scalar}

\author{Hieu The Pham$^{1,2}$}
\email{19130159@student.hcmus.edu.vn}
\author{Eibun Senaha$^{3,4}$}
\email{eibunsenaha@vlu.edu.vn (corresponding author)}
\affiliation{$^1$Department of Theoretical Physics, University of Science, Ho Chi Minh City, Vietnam}
\affiliation{$^2$Vietnam National University, Ho Chi Minh City, Vietnam}
\affiliation{$^3$Subatomic Physics Research Group, Science and Technology Advanced Institute, Van Lang University, Ho Chi Minh City, Vietnam}
\affiliation{$^4$Faculty of Applied Technology, School of Technology, Van Lang University, Ho Chi Minh City, Vietnam}

\date{\today}

\begin{abstract}
We study domain wall induced by spontaneously broken $\mathbb{Z}_2$ symmetry and its gravitational wave signature in the standard model with a complex scalar in connection with dark matter physics. 
In a minimal setup, a linear term of the singlet field is added to the scalar potential as an explicit $\mathbb{Z}_2$ breaking term to make the domain wall unstable. We obtain its minimal size from cosmological constraints and show that the parameter space that can be probed by current and future pulsar time array experiments requires the vacuum expectation value of the singlet field to be greater than $\mathcal{O}(10-100)$ TeV, along with a singlet-like Higgs mass of $\mathcal{O}(1-100)$ TeV. However, such a region is severely restricted by the dark matter relic density, which places an upper bound on the singlet vacuum expectation value at approximately 200 TeV, and limits the dark matter mass to about half of the singlet-like Higgs boson mass.
\end{abstract}


\maketitle


\section{Introduction}\label{sec:intro}
The search for physics beyond the standard model (SM) is the most critical and pressing issue in particle physics. 
After discovering the Higgs boson with a mass of 125 GeV~\cite{ATLAS:2012yve,CMS:2012qbp}, attention has increasingly been drawn to searches for extra scalar particles. However, despite current experimental data from the Large Hadron Collider (LHC), there is still no clear evidence of new physics. This may suggest that the new mass scales are too high to be reached, or that the new interactions are too weak to be detected. In such a situation, the use of cosmological probes through gravitational waves (GWs) can complement the search for extended Higgs sectors (for reviews, see, e.g., Refs.~\cite{Saikawa:2017hiv,Athron:2023xlk,Roshan:2024qnv}).

Currently, positive evidence of the stochastic gravitational wave background in the nano-Hz frequency range is reported by 
pulsar timing array experiment collaborations, NANOGrav~\cite{NANOGrav:2023gor,*NANOGrav:2023hde,*NANOGrav:2023ctt,*NANOGrav:2023hvm,*NANOGrav:2023hfp,*NANOGrav:2023tcn,*NANOGrav:2023icp}, EPTA~\cite{EPTA:2023fyk}, PPTA~\cite{Reardon:2023gzh}, and CPTA~\cite{Xu:2023wog}.
In addition to the supermassive black hole binaries interpretation~\cite{Bi:2023tib,*Ellis:2023dgf}, alternative explanations by extended Higgs sectors have also been  proposed~\cite{Wu:2023hsa,Ellis:2023oxs,Maji:2023fba,*Kitajima:2023cek,*Lazarides:2023ksx,*King:2023cgv,*Lu:2023mcz,*Barman:2023fad,*Zhang:2023nrs,Fujikura:2023lkn,*Li:2023bxy,*Xiao:2023dbb,*Chen:2023bms,Babichev:2023pbf}.

If there exist multiscalar fields, cosmological phase transitions would be diverse. For example, the electroweak phase transition (EWPT), which is a smooth crossover in the SM~\cite{Kajantie:1996mn,Rummukainen:1998as,Csikor:1998eu,Aoki:1999fi}, could be of first order, and GWs could be produced by bubble collisions, etc. 
This possibility is particularly important in the electroweak baryogenesis mechanism~\cite{Kuzmin:1985mm} (for reviews, see, e.g., Refs.\cite{Cohen:1993nk,Rubakov:1996vz,Funakubo:1996dw,Riotto:1998bt,Trodden:1998ym,Quiros:1999jp,Bernreuther:2002uj,Cline:2006ts,Morrissey:2012db,Konstandin:2013caa,Senaha:2020mop}). 
Additionally, other phase transitions prior to EWPT could occur, in which case extra sources of GWs could exist, such as collapses of domain walls (DWs) induced by spontaneously broken discrete symmetries. In such situations, the order of the phase transitions is not necessarily of first order.

One of the simplest new physics models is the SM with a complex scalar (cxSM)~\cite{Barger:2008jx,Gonderinger:2012rd,*Costa:2014qga,*Chiang:2017nmu,*Abe:2021nih,*Cho:2021itv,*Jiang:2015cwa,*Grzadkowski:2018nbc,*Cho:2022our,*Egle:2022wmq,*Egle:2023pbm,*Zhang:2023mnu,Barger:2010yn,Chen:2020wvu}.
Its scalar potential possesses a global U(1) symmetry if it is a function of $|\mathbb{S}|^2$, where $\mathbb{S}$ denotes a complex singlet scalar field. To avoid the emergence of a massless Nambu-Goldstone boson after the U(1) symmetry is spontaneously broken, we need to add explicit U(1) breaking terms, such as a $\mathbb{S}^2$ term, which breaks the U(1) symmetry down to a $\mathbb{Z}_2$ symmetry. In this case, the DW would appear if the $\mathbb{Z}_2$ is spontaneously broken. Adding a linear term in the $\mathbb{S}$ field to the scalar potential is a common way to avoid the cosmologically unwanted DW. 
In the CP conserving limit, the scalar potential becomes invariant under $\mathbb{S}\to \mathbb{S}^*$, or equivalently $\text{Im}\mathbb{S}\to-\text{Im}\mathbb{S}$.\footnote{DW induced by the CP symmetry (called CPDW) in the cxSM is extensively studied in Ref.~\cite{Chen:2020wvu}}.
Therefore, $\text{Im}\mathbb{S}$ can be dark matter (DM). As first indicated in Ref.~\cite{Barger:2010yn}, a spin-independent DM cross section with nucleons would vanish if the linear term is absent (for this type of cancellation mechanism, see also Ref.~\cite{Gross:2017dan}).
To our knowledge, the lower value of the biased term required by the DW collapse has not been explicitly quantified, taking the DM constraints into account.

In this article, we explore DW and its GW signatures in the CP-conserving cxSM with the minimal setup, where the U(1)-breaking terms in the scalar potential are only $\mathbb{S}$ and $\mathbb{S}^2$. After considering constraints such as vacuum stability, tree-level unitarity, the DM relic density, and the condition that DW decays before the big-bang nucleosynthesis (BBN) era, we obtain the lower bound of the linear term in $\mathbb{S}$ and identify a parameter space that is accessible by future experiments such as SKA~\cite{Janssen:2014dka}. In this work, we do not aim to explain the NANOGrav 15-year (NG15) data using the current model since the interpretation by DW with a constant tension, which applies to our case, is not favored~\cite{Babichev:2023pbf}. Instead, we use the NG15 data as a constraint when selecting our benchmark points.
Our analysis shows that the coefficient of the linear term needs to be higher than $\mathcal{O}(10^{-15})~\text{GeV}^3$, and in order to have the detectable GW spectrum, the vacuum expectation value (VEV) of $\mathbb{S}$ needs to be greater than $\mathcal{O}(10-100)$ TeV. Additionally, the mass of the singlet-like Higgs boson should be within the range of $\mathcal{O}(1-100)$ TeV with a mixing angle between the two Higgs bosons of approximately $\mathcal{O}(0.10-10)$ degrees.
Such mass ranges are similar to those in the CPDW case~\cite{Chen:2020wvu}. However, the allowed region in our scenario is severely limited by the DM relic abundance, imposing an upper limit of about 200 TeV on the singlet Higgs VEV and limiting the DM mass to about half the mass of the singlet-like Higgs boson.

The paper is organized as follows. In Section \ref{sec:model}, we introduce the cxSM and present the masses and couplings at the tree level. We also describe the theoretical constraints such as vacuum stability and perturbative unitarity in this section. In Section \ref{sec:dw}, we derive the equations of motion for DW and display DW profiles using a typical parameter set. We then discuss the decays of DWs in terms of the BBN constraint and GWs signatures, considering the discovery potential at SKA.
Our main numerical results are presented in Section \ref{sec:results}, and the conclusion is made in Section \ref{sec:conclusion}.

\section{Model}\label{sec:model}
The cxSM is an extension of the SM that involves adding a complex scalar field denoted as $\mathbb{S}$. The scalar potential of this model generally has 13 parameters. However, to simplify the model, the potential is modified by enforcing a global U(1), and some symmetry-breaking terms are added~\cite{Barger:2008jx}. In the simplest model, the scalar potential is defined as
\begin{align}
V_0(H, \mathbb{S}) = \frac{m^2}{2}H^\dagger H+\frac{\lambda}{4}(H^\dagger H)^2
	+\frac{\delta_2}{2}H^\dagger H|\mathbb{S}|^2+\frac{b_2}{2}|\mathbb{S}|^2+\frac{d_2}{4}|\mathbb{S}|^4
	+\bigg(a_1\mathbb{S}+\frac{b_1}{4}\mathbb{S}^2+{\rm H.c.}\bigg),
\end{align}
where $b_1$ is needed to avoid the unwanted massless Nambu-Goldstone boson associated with the spontaneously broken global U(1) symmetry. On the other hand, $a_1$ has a dual role: not only does it break the U(1) symmetry, but it also introduces a bias that destabilizes DW generated by the spontaneously broken $\mathbb{Z}_2$ symmetry, thereby avoiding cosmological constraints. As mentioned in Introduction, we quantify the required magnitude of $a_1$ in Sec.~\ref{sec:results}. Although $a_1$ and $b_1$ can be complex parameters, only their relative phase gives rise to the physical phase. However, for our present analysis, we assume them to be real, so that the scalar potential $V_0$ remains invariant under the CP transformation $\mathbb{S}\to \mathbb{S}^*$.

The scalar fields are parametrized as
\begin{align}
H(x) &=
	\left(
		\begin{array}{c}
		G^+(x) \\
		\frac{1}{\sqrt{2}}\big(v+h(x)+iG^0(x)\big)
		\end{array}
	\right),\\
\mathbb{S}(x) &= \frac{1}{\sqrt{2}}\big(v_S+ S(x)+i\chi(x) \big),
\end{align}
where $v(\simeq 246.22~\text{GeV}$) and $v_S$ are the VEVs of the doublet and singlet scalar fields, respectively. $G^{+,0}(x)$ are the unphysical Nambu-Goldstone bosons associated with electroweak symmetry breaking. The CP-even scalars $h(x)$ and $S(x)$ can mix and one of them becomes the SM-like Higgs boson with a mass of 125 GeV. One can see that the scalar potential $V_0$ is invariant under $\chi(x)\to -\chi(x)$ due to the aforementioned CP invariance, which implies that $\chi(x)$ can play a role of DM.

The tadpole conditions with respect to $h$ and $S$ are, respectively, given by
\begin{align}
\left\langle \frac{\partial V_0}{\partial h}\right\rangle & = 
v\left[
\frac{m^2}{2} + \frac{\lambda}{4}v^2+\frac{\delta_2}{4}v_S^2 
\right]= 0,\label{tad_h}\\
\left\langle \frac{\partial V_0}{\partial S}\right\rangle & = 
v_S
\left[
\frac{b_1+b_2}{2}+\frac{\delta_2}{4}v^2+\frac{d_2}{4}v_S^2+\frac{\sqrt{2}a_1}{v_S}
\right]
 = 0,\label{tad_s}
\end{align}
where $\langle \cdots \rangle$ denotes that the fluctuation fields are taken zero after the derivatives.

The mass matrix of the $(h,S)$ bosons takes the form
\begin{align}
\mathcal{M}_S^2 &= 
\left(
	\begin{array}{cc}
	\frac{m^2}{2}+\frac{3\lambda}{4}v^2+\frac{\delta_2}{4}v_{S}^2 &
	\frac{\delta_2}{2}vv_{S} \\
	\frac{\delta_2}{2}vv_{S} & \frac{b_1+b_2}{2}+\frac{\delta_2}{4}v^2+\frac{3d_2}{4}v_{S}^2
	\end{array}
\right) = 
\left(
\begin{array}{cc}
	\frac{\lambda}{2}v^2 & \frac{\delta_2}{2}vv_{S} \\
	\frac{\delta_2}{2}vv_{S} & \frac{d_2}{2}v_{S}^2-\frac{\sqrt{2}a_1}{v_{S}}
\end{array}
\right),\label{Ms_tree}
\end{align}
where the tadpole conditions (\ref{tad_h}) and (\ref{tad_s}) are used in the second equality.
This mass matrix can be diagonalized by an orthogonal matrix $O(\alpha)$
\begin{align}
O^T(\alpha)\mathcal{M}_S^2O(\alpha)  =
\left(
	\begin{array}{cc}
	m_{h_1}^2 & 0 \\
	0 & m_{h_2}^2
	\end{array}
\right), \quad
O(\alpha) = 
\left(
	\begin{array}{cc}
	c_\alpha & -s_\alpha \\
	s_\alpha & c_\alpha
	\end{array}
\right),
\end{align}
where $s_\alpha=\sin\alpha$ and $c_\alpha=\cos\alpha$ with $-\pi/4\le\alpha\le\pi/4$ and $h_1$ is identified as the SM-like Higgs boson in our study, i.e., $m_{h_1}=125$ GeV.
In the large $v_{S}$ limit, the two scalar masses are simplified to
\begin{align}
m_{h_1}^2 \simeq \frac{1}{2}\left(\lambda-\frac{\delta_2^2}{d_2}\right)v^2, \quad
m_{h_2}^2 \simeq \frac{1}{2}\left(d_2 v_{S}^2+\frac{\delta_2^2}{d_2}v^2\right).
\end{align}
The DM mass is given by  
\begin{align}
m_\chi^2 = \frac{b_2-b_1}{2}
	+\frac{\delta_2}{4}v^2+\frac{d_2}{4}v_{S}^2
	= -\frac{\sqrt{2}a_1}{v_{S}}-b_1, \label{mDM_tree}
\end{align}
where the tadpole condition (\ref{tad_s}) is used in the second equality.

There are 7 parameters in the scalar potential: $(m^2, b_2, \lambda, d_2, \delta_2, b_1, a_1)$. All parameters except for $a_1$ can be expressed in terms of $(v, v_S, m_{h_1}, m_{h_2}, \alpha, m_\chi)$. More explicitly, one finds
\begin{align}
\lambda 
&= \frac{2}{v^2}(m_{h_1}^2c_\alpha^2+m_{h_2}^2s_\alpha^2), \label{lam}\\
d_2 &= \frac{2}{v_{S}^2}
\left[
	m_{h_1}^2s_\alpha^2+m_{h_2}^2c_\alpha^2+\frac{\sqrt{2}a_1}{v_{S}}
\right],\label{d2} \\
\delta_2 &= \frac{s_{2\alpha}}{vv_{S}}(m_{h_1}^2-m_{h_2}^2), \label{del2}\\
b_1 &= -\left[m_\chi^2+\frac{\sqrt{2}a_1}{v_{S}}\right], \\
m^2 &= -\frac{\lambda}{2}v^2-\frac{\delta_2}{2}v_{S}^2,\label{tad_msq} \\
b_2 &= -\frac{\delta_2}{2}v^2-\frac{d_2}{2}v_{S}^2
	-\frac{2\sqrt{2}a_1}{v_{S}}-b_1.\label {tad_b2}
\end{align}
The Higgs couplings to fermions and gauge bosons are, respectively, given by
\begin{align}
\mathcal{L}_{h_i\bar{f}f} &= -\frac{m_f}{v}\sum_{i=1,2}\kappa_{if} h_i\bar{f}f, \\
\mathcal{L}_{h_iVV} &= \frac{1}{v}\sum_{i=1,2}\kappa_{iV}h_i(m_Z^2Z_\mu Z^\mu+2m_W^2W_\mu^+W^{-\mu}),
\end{align}
 where $\kappa_{1f}=\kappa_{1V}=c_\alpha$ and $\kappa_{2f}=\kappa_{2V}=-s_\alpha$. 
The values of $m_{h_2}$ and $\kappa$ are restricted by LHC data~\cite{ATLAS:2022vkf,CMS:2022dwd}.
Regarding theoretical bounds, on the other hand, we impose the bounded-from-below condition and the global minimum condition on the scalar potential.
The former is given by 
\begin{align}
\lambda>0,\quad d_2>0,\quad -\sqrt{\lambda d_2}<\delta_2,
\end{align}
where the third inequality condition is needed only if $\delta_2<0$. 
The global minimum condition means that $V_0(v,v_S)$ is smaller than any other potential energy. 
Moreover, we impose the perturbative unitarity~\cite{Chen:2020wvu}. 
Because of this, the value of $m_{h_2}$ has an upper bound for a given $\alpha$.
In the parameter space where the sizable GW is produced, the theoretical constraints hold more significance than the collider bounds, as discussed in Sec.~\ref{sec:results}.

\section{Collapose of Domain wall and its gravitatioal wave signatures}\label{sec:dw}
In the case of $a_1=0$ and $v_S\neq0$, we have the DW solution. 
The classical scalar fields are parameterized as\footnote{Since we consider the CP-conserving case, only the real part of the singlet field would be relevant.}
\begin{align}
\langle H(z)\rangle=\frac{1}{\sqrt{2}}
\left(
	\begin{array}{c}
		0 \\
		\phi(z)
	\end{array}
\right),\quad
\langle S(z)\rangle=\frac{\phi_S(z)}{\sqrt{2}}, \label{dw_fields}
\end{align}
where $z$ is a coordinate perpendicular to DW. The energy density of DW is given by
\begin{align}
\mathcal{E}_{\text{DW}} =
	\frac{1}{2}(\partial_z \phi)^2+\frac{1}{2}(\partial_z \phi_S)^2+V(\phi, \phi_S),
\end{align}
with the normalized scalar potential
\begin{align}
V(\phi,\phi_S) &= V_0(\phi, \phi_S)-V_0(v, v_S) \nonumber \\
&=  \frac{\lambda}{16}(\phi^2-v^2)^2
	+\frac{\delta_2}{8}(\phi^2-v^2)(\phi_S^2-v_S^2)+\frac{d_2}{16}(\phi_S^2-v_S^2)^2,
\end{align}
where Eqs.~(\ref{tad_msq}) and (\ref{tad_b2}) are used to eliminate $m^2$ and $b_1+b_2$.
The subtraction of $V_0(v,v_S)$ is necessary in order not to generate a divergence in a tension of DW defined below. 
From Eqs.~(\ref{lam})-(\ref{del2}) with $a_1=0$, we note that $V(\phi,\phi_S)$ is determined by $m_{h_2}$, $\alpha$, and $v_S$, independent of $m_\chi$.

The equations of motion for DW are
\begin{align}
\frac{d^2\Phi}{dz^2}-\frac{\partial V}{\partial \Phi}&=0, \label{eom_dw}
\end{align}
where $\Phi=\{\phi, \phi_S\}$ with the boundary conditions
\begin{align}
\lim_{z\to\pm\infty}\phi(z)&=v,\quad \lim_{z\to\pm\infty}\phi_S(z)=\pm v_S.\label{eom_dw_bc}
\end{align}
We solve Eqs.~(\ref{eom_dw}) and (\ref{eom_dw_bc}) using a \textit{relaxation method}~\cite{Press:1992zz}, and then calculate the tension of DW which is given by
\begin{align}
\sigma_{\text{DW}} = \int_{-\infty}^\infty dz~\mathcal{E}_{\text{DW}} = \sigma_{\text{DW}}^{\text{kin}}+\sigma_{\text{DW}}^{\text{pot}},
\end{align}
where $\sigma_{\text{DW}}^{\text{kin}}$ and $\sigma_{\text{DW}}^{\text{pot}}$ are contributions of kinetic and potential terms, respectively. From Derrick's theorem~\cite{Derrick:1964ww}, it follows that $\sigma_{\text{DW}}^{\text{kin}}=\sigma_{\text{DW}}^{\text{pot}}$. We use this relation as a cross-check for the correctness of our numerical solutions.

In the decoupling limit, where $|\alpha| \ll 1$, DW is reduced to that in the $\phi^4$ theory (see, e.g., Ref.~\cite{Kolb:1990vq}). 
In this case, the DW profile is expressed as
\begin{align}
\phi_S(z) = v_S\tanh\left(\sqrt{\frac{d_2}{8}}v_Sz \right).
\end{align}
Using this analytic solution, $\sigma_{\text{DW}}$ can be easily caculated as
\begin{align}
\sigma_{\text{DW}} 
&\simeq \frac{d_2}{8}v_S^3\int_{-\infty}^\infty d\xi~\left[\tanh^2\left(\sqrt{\frac{d_2}{8}}\xi\right)-1 \right]^2
= \frac{2}{3}\sqrt{\frac{d_2}{2}}v_S^3 \simeq \frac{2}{3}m_{h_2}v_S^2,
\label{sigma_dw_app}
\end{align}
where $\xi = v_Sz$, and Eq.~(\ref{d2}) is used. Therefore, the magnitude of $\sigma_{\text{DW}}$ is controlled by $m_{h_2}v_S^2$ in the decoupling limit.

\begin{figure}[t]
\center
\includegraphics[width=7cm]{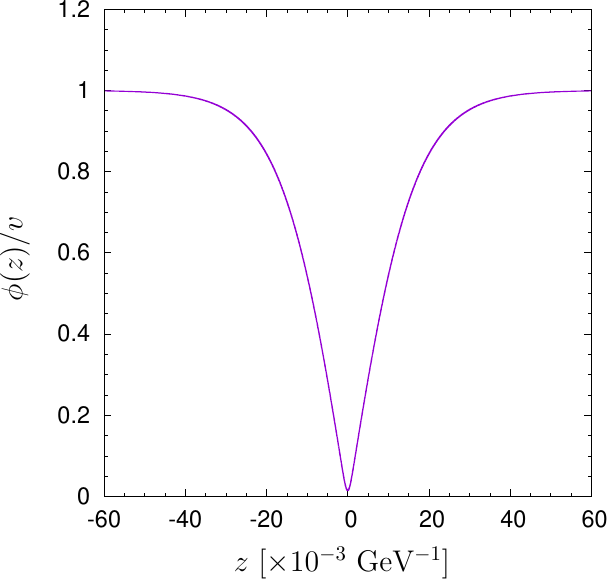}
\hspace{0.5cm}
\includegraphics[width=7cm]{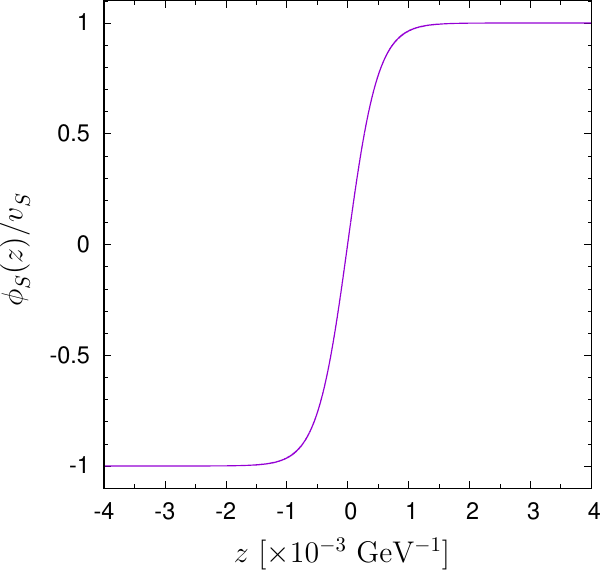}
\caption{The DW profiles of $\phi(z)$ (left) and $\phi_S(z)$ (right), respectivley. We take $m_{h_2}=4.0$ TeV, $m_\chi=2.0$ TeV, $v_S = 100$ TeV, and $\alpha=0.10^\circ~(=1.7\times 10^{-3}~\text{radians})$. This parameter set gives $\sigma_{\text{DW}}=2.7\times 10^{13}~[\text{GeV}^3]$.}
\label{fig:dwprofs}
\end{figure}

Now, we present a DW solution in a parameter space of our interest, where $m_{h_2}=4.0$ TeV, $m_\chi=2.0$ TeV, $v_S = 100$ TeV, and $\alpha=0.10^\circ~(=1.7\times 10^{-3}~\text{radians})$. This choice turns out to be consistent with all theoretical and experimental constraints and yields detectable GW signals in future experiments discussed below.
The left and right panels of Fig.~\ref{fig:dwprofs} show the DW profiles of $\phi(z)$ and $\phi_S(z)$, respectively.
$\phi_S(z)$ has a hyperbolic tangent shape as expected, while $\phi(z)$ has a dip around $z\simeq 0$, which should be attributed to the singlet-doublet scalar field mixing. 
In this parameter set, we obtain $\sigma_{\text{DW}}=2.7\times 10^{13}~[\text{GeV}^3]$, which agrees well with the approximate expression~(\ref{sigma_dw_app}).

DW must disappear to avoid interfering with cosmological observations, and the $\mathbb{Z}_2$ breaking term $a_1\mathbb{S}$ is introduced to destabilize the DW. In the presence of nonzero $a_1$, the degeneracy of the two vacua is broken by 
\begin{align}
\Delta V \equiv |V(v,v_S)-V(v,-v_S)| = 2\sqrt{2}|a_1|v_S.
\end{align}
DW would be annihilated when the pressure of DW is less than the pressure caused by the bias term $\Delta V$.
We determine the annihilation temperature by the condition $\Delta V = C_{\text{ann}}\frac{\mathcal{A}\sigma_{\text{DW}}}{t_{\text{ann}}}$~\cite{Saikawa:2017hiv}, where $C_{\text{ann}} \simeq 2-5$~\cite{Saikawa:2017hiv} and $\mathcal{A}\simeq 0.8\pm0.1$~\cite{Hiramatsu:2013qaa}. For our numerical analysis, we assume that $C_{\text{ann}}=2$ and $\mathcal{A}=0.8$. 
For convenience, we introduce the dimensionless DW tension
\begin{align}
\hat{\sigma}_{\text{DW}} = \frac{\sigma_{\text{DW}}}{m_{h_2}v_S^2}.
\end{align}
The successful BBN enforces a bound on $t_{\text{ann}} = C_{\text{ann}}\mathcal{A}\frac{\sigma_{\text{DW}}}{\Delta V}<t_{\text{BBN}}\equiv 0.01$ s, which places a lower bound on $|a_1|$ as
\begin{align}
|a_1|&> 2.3\times 10^{-15}~\text{GeV}^3\left(\frac{m_{h_2}}{10^3~\text{GeV}}\right) \left(\frac{v_S}{10^5~\text{GeV}}\right)C_{\text{ann}}\mathcal{A}\hat{\sigma}_{\text{DW}}.
\label{a1_low_BBN}
\end{align}
For the parameter set considered in Fig.~\ref{fig:dwprofs}, $|a_1|$ should be greater than $\mathcal{O}(10^{-15})~\text{GeV}^3$ for the sucessful BBN. 
If DW is annihilated when the universe is in the radiation-dominated era, the temperature at $t=t_{\text{ann}}$ is given by
\begin{align}
T_{\text{ann}}
&= 1.8\times 10^{-2}~\text{GeV}\left(\frac{g_*(T_{\text{ann}})}{10}\right)^{-1/4} 
\left(\frac{|a_1|}{10^{-15}~\text{GeV}^3}\right)^{1/2}\left(\frac{10^3~\text{GeV}}{m_{h_2}}\right)^{1/2}\left(\frac{10^5~\text{GeV}}{v_S}\right)^{1/2}\nonumber \\
&\hspace{1cm}\times  (C_{\text{ann}} \mathcal{A} \hat{\sigma}_{\text{DW}} )^{-1/2},
\end{align}
which has to be greater than $T_{\text{BBN}}=8.6\times 10^{-3}$ GeV. Note that the larger $|a_1|$ makes $T_{\text{ann}}$ higher. 

We consider another constraint that the DWs should not dominate the universe. Assuming that the energy density of the universe is initially dominated by radiations, then the time of the DW-dominated universe can be calculated as $t_{\text{dom}}=3m_{\text{pl}}^2/(32\pi\mathcal{A}\sigma_{\text{DW}})$~\cite{Saikawa:2017hiv}, where $m_{\text{pl}}=1.22\times 10^{19}$ GeV. From the condition $t_{\text{ann}}<t_{\text{dom}}$, it follows that
\begin{align}
|a_1|>8.0\times 10^{-18}~\text{GeV}^3\left(\frac{m_{h_2}}{10^3~\text{GeV}}\right)^2\left(\frac{v_S}{10^5~\text{GeV}}\right)^3C_{\text{ann}}\mathcal{A}^2\hat{\sigma}_{\text{DW}}^2.
\label{a1_low_DW}
\end{align}
This constraint is weaker than that in Eq.~(\ref{a1_low_BBN}) for the parameter set used in Fig.~\ref{fig:dwprofs}. 
In principle, however, it could impose a more stringent bound in the region of larger $m_{h_2}$ and/or $v_S$ due to their higher powers.

After the annihilation of DWs, GW would be generated, and their spectrum at peak frequency can be determined by~\cite{Hiramatsu:2012sc,*Kawasaki:2014sqa,Saikawa:2017hiv}\footnote{A recent study by~Ref.~\cite{Kitajima:2023kzu} has provided somewhat different estimates. However, our main conclusion remains qualitatively unchanged.}
\begin{align}
f_{\text{peak}} &= 1.1\times 10^{-9}~\text{Hz}\left(\frac{g_*(T_{\text{ann}})}{10} \right)^{1/2}
\left(\frac{g_{*s}(T_{\text{ann}})}{10}\right)^{-1/3}\left(\frac{T_{\text{ann}}}{10^{-2}~\text{GeV}} \right),\\
\Omega_{\text{GW}}h^2(f_{\text{peak}}) 
&= 7.2\times 10^{-10}~\tilde{\epsilon}_{\text{GW}}\mathcal{A}^2
\left(\frac{g_{*s}(T_{\text{ann}})}{10}\right)^{-4/3}\left(\frac{T_{\text{ann}}}{10^{-2}~\text{GeV}} \right)^{-4} \nonumber\\
&\quad \times 
\left(\frac{m_{h_2}}{10^3~\text{GeV}}\right)^2\left(\frac{v_S}{10^5~\text{GeV}}\right)^4 \hat{\sigma}_{\text{DW}}^2,
\label{OmegGW}
\end{align}
where $\tilde{\epsilon}_{\text{GW}}=0.7\pm 0.4$~\cite{Hiramatsu:2013qaa}, $g_*(T_{\text{ann}})=g_{*s}(T_{\text{ann}})=10.75$ for $1~\text{MeV}<T_{\text{ann}}\ll 100~\text{MeV}$~\cite{Kolb:1990vq}.
As seen from Eq.~(\ref{OmegGW}), $\Omega_{\text{GW}}h^2(f_{\text{peak}})$ is proportional to $\sigma^2_{\text{DW}}$ and can reach $\mathcal{O}(10^{-10})$ for the parameter set in Fig.~\ref{fig:dwprofs}.
For an arbitrary frequency $f$, we use
\begin{align}
\Omega_{\text{GW}}h^2(f<f_{\text{peak}}) &= \Omega_{\text{GW}}h^2(f_{\text{peak}})\left(\frac{f}{f_{\text{peak}}}\right)^3, \\
\Omega_{\text{GW}}h^2(f>f_{\text{peak}}) &= \Omega_{\text{GW}}h^2(f_{\text{peak}})\left(\frac{f_{\text{peak}}}{f}\right).
\end{align}
The signal-to-noise ratio (SNR) is defined as
\begin{align}
\text{SNR} = \sqrt{t_{\text{dur}}\int_{f_{\text{min}}}^{f_{\text{max}}}df\left(\frac{\Omega_{\text{GW}}(f)h^2}{\Omega_{\text{exp}}(f)h^2} \right)^2 },\label{SNR}
\end{align}
where $t_{\text{dur}}$ denotes the duration of the mission. In our numerical study, we explore the parameter space assuming $t_\text{dur}=20$ and $\text{SNR}=20$ at the SKA experiment~\cite{Janssen:2014dka} which aims at detecting GW in the nano-Hz frequency. 
Such a region is currently favored by the pulsar timing array experimental data~\cite{NANOGrav:2023gor,*NANOGrav:2023hde,*NANOGrav:2023ctt,*NANOGrav:2023hvm,*NANOGrav:2023hfp,*NANOGrav:2023tcn,*NANOGrav:2023icp,EPTA:2023fyk,Reardon:2023gzh,Xu:2023wog}.
The requirement of the GW signal detectability puts an upper bound on $|a_1|$ because of $\Omega_{\text{GW}}\propto T_{\text{ann}}^{-4}\propto |a_1|^{-2}$, which has to be consistent with the lower bound on $|a_1|$ coming from either Eq.~(\ref{a1_low_BBN}) or Eq.~(\ref{a1_low_DW}).

\section{Dark matter}\label{sec:dm}
\begin{figure}[t]
\center
\includegraphics[width=16cm]{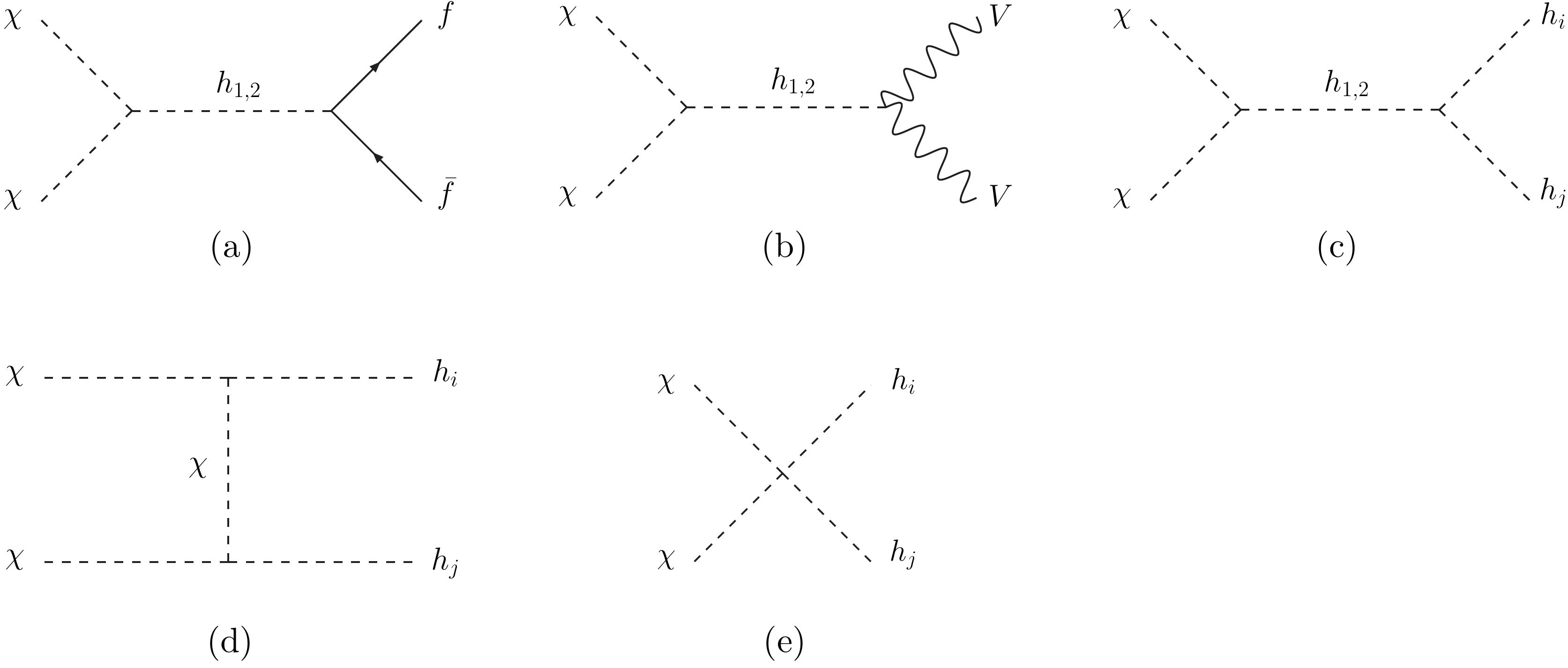}
\caption{The DM annihilation processes, where $f$ denote the SM fermions, while $V$ represent $W^\pm$ and $Z$. Each process is described by Eqs.~(\ref{sigmav_a})-(\ref{sigmav_e}) in Appendix~\ref{app:dm}.}
\label{fig:DM}
\end{figure}
The parameter space of the CP-conserving cxSM is further constrained by DM physics. Fig.~\ref{fig:DM} shows the annihilation processes of $\chi$, and each contribution is given by Eqs.~(\ref{sigmav_a})-(\ref{sigmav_e}) in Appendix \ref{app:dm}.
Let us consider the cases where $\Omega_{\text{GW}}(\propto \sigma_{\text{DW}}^2 \propto m_{h_2}^2v_S^4)$ is enhanced and the mixing angle $\alpha$ is small enough to make $h_1$ similar to the SM. In this scenario, $\delta_2$ would be small (as per Eq.(\ref{del2})), and some Higgs couplings appearing in the DM annihilation processes can become small as well. This would lead to a significant suppression of the DM annihilation cross section, which could cause the DM relic density (denoted as $\Omega_\chi$) to exceed the observed value of $\Omega_{\text{DM}}h^2=0.11933\pm 0.00091$~\cite{Planck:2018vyg}. However, such an overabundance of DM can be avoided in the resonance region, where the DM mass is half as large as the masses of the intermediate Higgs bosons.
Among the processes given in Fig.~\ref{fig:DM}, the most relevant diagram in the case of TeV DM mass would be diagram (c), which is cast into the form
\begin{align}
(\sigma v_{\text{rel}})_{\chi\chi\to h_{2}\to h_ih_j}
& \simeq \mathcal{S}\frac{\beta_{h_ih_j}}{8\pi s}
\left|
\frac{\lambda_{h_1\chi\chi}\lambda_{h_1h_ih_j}}{s-m_{h_1}^2+im_{h_1}\Gamma_{h_1}}
+\frac{\lambda_{h_2\chi\chi}\lambda_{h_2h_ih_j}}{s-m_{h_2}^2+im_{h_2}\Gamma_{h_2}}
\right|^2,
\end{align}
where $\mathcal{S}=1/2$ for identical particles as final states, and $\mathcal{S}=1$ otherwise. The center-of-mass energy squared is approximately given by $s\simeq 4m_\chi^2$, and the phase space factor is
\begin{align}
\beta_{ij} = 
\sqrt{1-\frac{2(m^2_{i}+m^2_{j})}{s}+\frac{(m^2_{i}-m^2_{j})^2}{s^2}}.
\end{align}
The Higgs couplings are explicitly given by Eqs.~(\ref{lam133})-(\ref{lam222}), and $\Gamma_{h_1}$ and $\Gamma_{h_2}$ are given by Eqs.~(\ref{Gam1}) and (\ref{Gam2}), respectively.
In the case of $s\simeq 4m_\chi^2 \simeq m_{h_2}^2\gg m_{h_1}^2$, for instance, $(\sigma v_{\text{rel}})_{\chi\chi\to h_{2}\to h_1h_1}$  would be reduced to
\begin{align}
(\sigma v_{\text{rel}})_{\chi\chi\to h_{2}\to h_1h_1}
& \simeq \frac{1} {16\pi m_{h_2}^2}\sqrt{1-\frac{4m_{h_1}^2}{m_{h_2}^2}}
\left|
\frac{\lambda_{h_1\chi\chi}\lambda_{h_1h_1h_1}}{m_{h_2}^2-m_{h_1}^2+ im_{h_2}\Gamma_{h_2}}
+\frac{\lambda_{h_2\chi\chi}\lambda_{h_2h_1h_1}}{im_{h_2}\Gamma_{h_2}}
\right|^2.
\end{align}
Note that the first term is suppressed by $\lambda_{h_1\chi\chi}\lambda_{h_1h_1h_1}\simeq (m_{h_1}^2s_\alpha/v_S)\cdot (3m_{h_1}^2c_\alpha^3/v)$ in the limit of $|a_1|/v_S^3\ll 1$. In the second term, on the other hand, the suppressed couplings $\lambda_{h_2\chi\chi}\lambda_{h_2h_1h_1}\simeq (m_{h_2}^2c_\alpha/v_S)\cdot (-(2 m_{h_1}^2 + m_{h_2}^2)c_\alpha^2 s_\alpha/v)$ could be compensated by the smallness of $\Gamma_{h_2}\propto s_\alpha^2$, preventing the annihilation cross section from becoming tiny.
From this simple argument, we expect the detectable GW region to require $m_{h_2}\simeq 2m_{\chi}$, otherwise, the DM would be overabundant and ruled out. We numerically verify this statement below. 

Now we move on to discuss the spin-independent (SI) cross section of DM scatting off a nucleon (denoted as $\sigma_{\text{SI}}^{N=p,n}$).   
After integrating the $h_i$ fields out, one can obtain an effective Lagrangian for DM-quarks interactions as
\begin{align}
\mathcal{L}_{\chi^2 \bar{q}q}
&= \chi^2\sum_q f_qm_q\bar{q}q,\quad f_q = \frac{1}{2v}
\left( 
	\frac{\lambda_{h_1\chi\chi}\kappa_{1q}}{m_{h_1}^2}
	+\frac{\lambda_{h_2\chi\chi}\kappa_{2q}}{m_{h_2}^2}
\right).
\end{align}
Using the coupling $f_q$, one can find $\sigma_{\text{SI}}^N$ as~\cite{Barger:2010yn,Gonderinger:2012rd}
\begin{align}
\sigma_{\rm SI}^N
&= \frac{1}{8\pi v^2}\frac{m_N^4}{(m_\chi+m_N)^2}
\frac{s_{2\alpha}^2(m_{h_1}^2-m_{h_2}^2)^2a_1^2}{m_{h_1}^4m_{h_2}^4v_S^4}
\left|
	\sum_{q=u,d,s} f_{T_q}+\frac{2}{9}f_{T_G}
\right|^2.
\end{align}
Currently, the value of $\sigma_{\text{SI}}^N$ is severely limited by the LZ experiment~\cite{LZ:2022lsv}. However, one can observe that $\sigma_{\text{SI}}^N$ would be zero if the value of $a_1$ were to be zero as well. This was first noted in Ref.~\cite{Barger:2010yn} and explored in greater detail in Ref.~\cite{Gross:2017dan}. Nonetheless, to avoid the DW problem, $a_1$ should not be exactly zero, as discussed in Sec.~\ref{sec:dw}.
In our study, we use \textsc{MicrOmegas}~\cite{Belanger:2006is,*Belanger:2008sj} to calculate $\Omega_\chi h^2$ and $\sigma_{\text{SI}}^p$. 

\begin{figure}[t]
\center
\includegraphics[width=7.5cm]{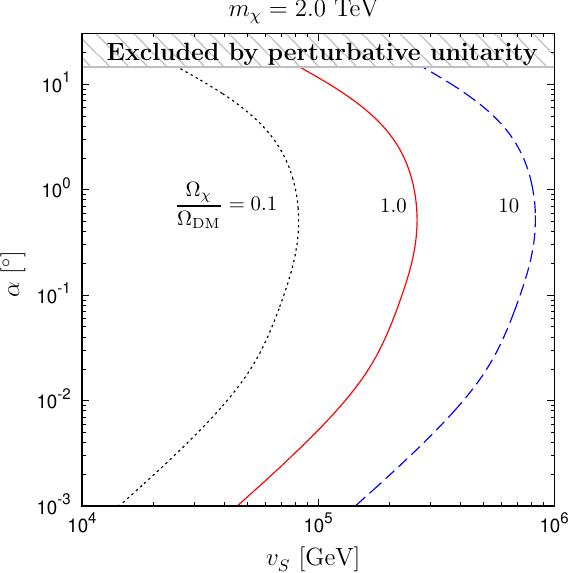} \\[1cm]
\includegraphics[width=7.5cm]{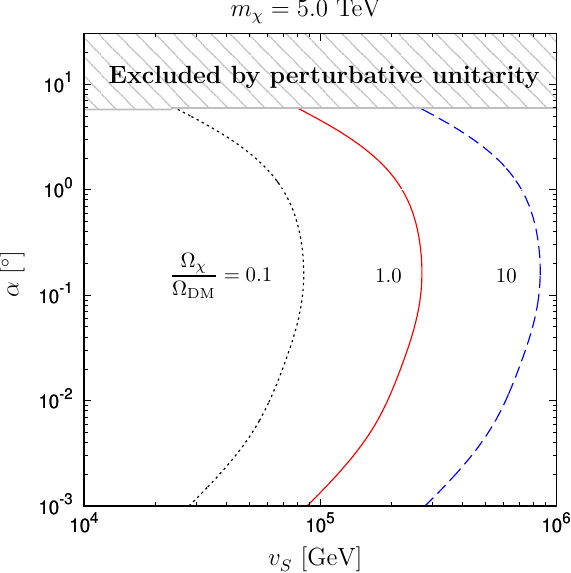} 
\hspace{0.5cm}
\includegraphics[width=7.5cm]{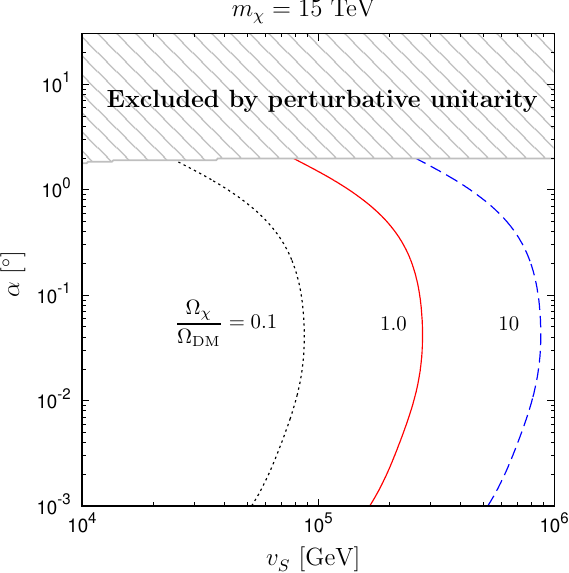} 
\caption{Contours of the DM relic abundance in the three cases: $m_\chi = 2.0$ TeV (upper panel), 5.0 TeV (lower-left panel), and 15 TeV (lower-right panel), respectively. In each panel, the three lines denote $\Omega_\chi/\Omega_\text{DM}=$0.1 (black, dotted line), 1.0 (red, solid line), and 10 (blue, dashed line), with $\Omega_\text{DM}$ representing the observed value of the DM relic abundance. The perturbative unitarity constraint excludes the shaded region. The second CP-even Higgs mass is set to $m_{h_2}=2m_\chi$ to enhance the DM annihilation crosse sections , while $a_1 = -10^{-12}~\text{GeV}^3$ for illustration.}
 \label{fig:DM_vS_alp}
\end{figure}

We close this section by giving the DM relic abundance in the region, where $v_S>10$ TeV, taking $m_\chi = m_{h_2}/2$.
Fig.~\ref{fig:DM_vS_alp} displays the contours of the DM relic abundance normalized by the observed value in three different DM mass cases: $m_\chi = 2.0$ TeV (upper panel), 5.0 TeV (lower-left panel), and 15 TeV (lower-right panel). The three lines in each panel represent $\Omega_\chi/\Omega_\text{DM}=$0.1 (black, dotted line), 1.0 (red, solid line), and 10 (blue, dashed line), respectively. The hatched area is excluded by the perturbative unitarity constraint. As an example, we take $a_1 = -10^{-12}~\text{GeV}^3$.
In all cases, we have $v_S\lesssim 200$ TeV in order not to exceed the observed DM relic abundance, implying that the value of $\sigma_{\text{DW}}(\propto v_S^2)$ would be bounded from above, which in turn can limit the magnitude of the GW spectrum.

\section{Numerical results}\label{sec:results}

\begin{figure}[t]
\center
\includegraphics[width=7.5cm]{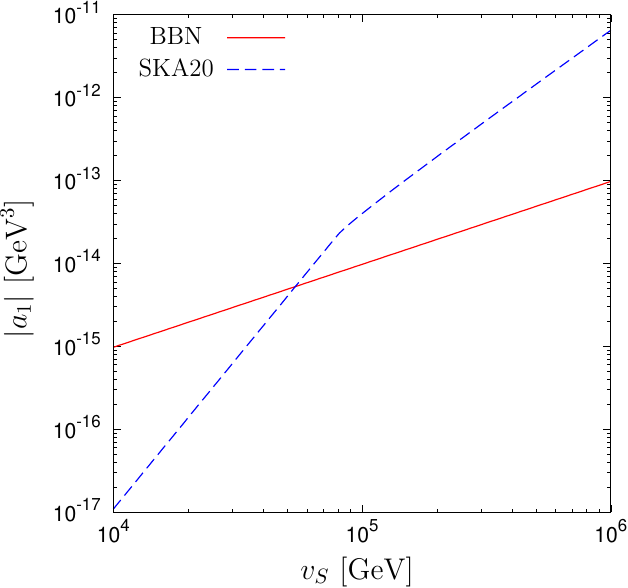}
\hspace{0.5cm}
\includegraphics[width=7.5cm]{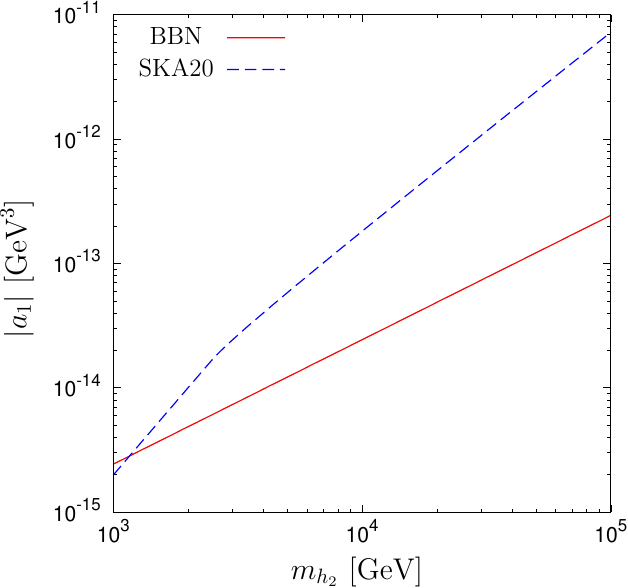}
\caption{Constraints on the biased term $|a_1|$ as a function of $v_S$ (left panel) and $m_{h_2}$ (right panel). We take $m_{h_2} = 4.0$ TeV in the left panel and $v_S=100$ TeV in the right panel, repspectively, and the remaining parameters are $m_{h_1} = 125$ GeV, $m_\chi = 2.0$ TeV, and $\alpha=0.10^\circ$.
The solid line in red represents the BBN bound which yields the lower bound on $|a_1|$. On the other hand, the dashed line in blue (SKA20) denotes the discovery potential case with SNR = 20 which sets the upper bound on $|a_1|$. 
}
\label{fig:GW}
\end{figure}

Here, we study the detectability of GW from the domain collapses. 
In Fig.~\ref{fig:GW}, we plot the constraints on the biased term $|a_1|$ as a function of $v_S$ (left panel) or $m_{h_2}$ (right panel) for  $m_{h_1} = 125$ GeV, $m_\chi = 2.0$ TeV, and $\alpha = 0.10^\circ$. In the left panel, we take $m_{h_2} = 4.0$ TeV, while for the right panel, $v_S=100$ TeV.
The red solid line represents the BBN bound, which sets the lower bound on $|a_1|$.
On the other hand, the blue-dashed line (denoted as SKA20) corresponds to the discovery potential at SKA with $\text{SNR}=20$, and above which $\text{SNR}<20$, yielding the upper bound on $|a_1|$. From the left panel, $v_S\gtrsim 54$ TeV is required to detect GW signals at SKA in the case of $m_{h_2}=4.0$ TeV. Similarly, from the right panel, $m_{h_2}\gtrsim 1.2$ TeV is necessary for the detectable GW signals for $v_S=100$ TeV. 
Therefore, the biased term has to satisfy that $|a_1|\gtrsim \mathcal{O}(10^{-15})~\text{GeV}^3$.
Due to the smallness of $a_1^2/v_S^4$ as well as $s_{2\alpha}\ll 1$, $\sigma_{\text{SI}}^N$ is far too small to be constrained by the current LZ data.
In the parameter space that we explore below, we always choose the value of $|a_1|$ so that it satisfies the condition of $\text{SNR} = 20$, while imposing the BBN constraint.

\begin{figure}[t]
\center
\includegraphics[width=7.5cm]{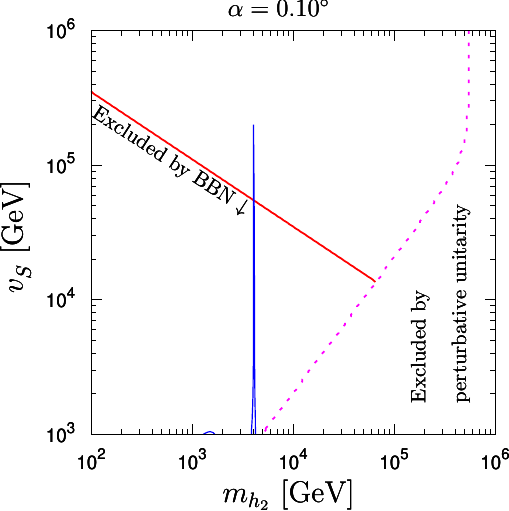} \\[0.5cm]
\includegraphics[width=7.5cm]{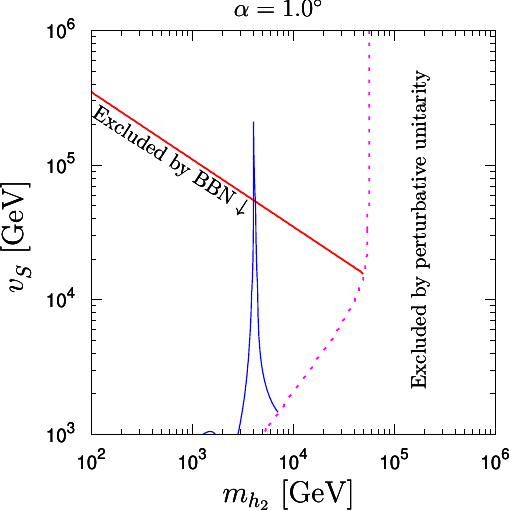} 
\hspace{0.5cm}
\includegraphics[width=7.5cm]{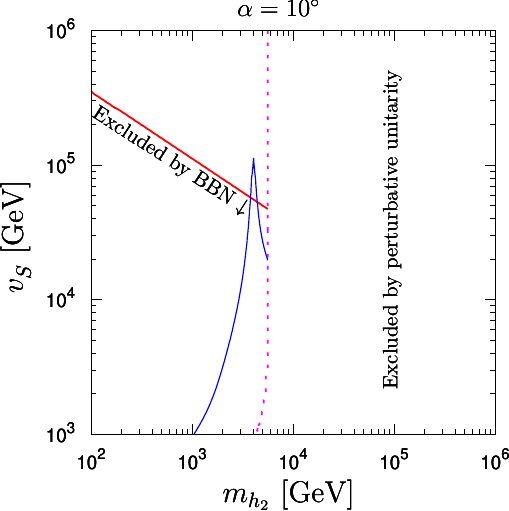}
\caption{Discovery potential at SKA as a function of $m_{h_2}$ and $v_S$. The lower region of the solid line is excluded by the BBN bound. The perturbative unitarity excludes the right region of the dotted line in magenta. The solid curve in blue shows the observed DM relic density $\Omega_\chi h^2=0.12$, and the narrower region rounded by the curve, $\Omega_\chi h^2<0.12$. 
Here, $\alpha=0.10^\circ$, $1.0^\circ$, and $10^\circ$ are taken in the top, bottom-left, and bottom-right panels, respectively.
The DM mass is fixed to $m_{\chi}=2.0$ TeV.
 }
 \label{fig:GW_mh2_vS}
\end{figure}

Fig.~\ref{fig:GW_mh2_vS} shows the discovery potential at SKA in the $(m_{h_2}, v_S)$ plane, where all the points satisfy $\text{SNR}=20$ by judiciously choosing $a_1$. We take $\alpha=0.10^\circ$ (upper panel),  and $\alpha = 1.0^\circ$ (lower-left panel), and $\alpha=10^\circ$ (lower-right panel), repecetively, and set $m_\chi = 2.0$ TeV.
The BBN constraint excludes the lower region of the solid line in red, while the right area of the dotted line in magenta is excluded by the perturbative unitarity. The solid curve in blue shows $\Omega_\chi h^2=0.12$, and the narrower region rounded by the curve corresponds to $\Omega_\chi h^2<0.12$. The allowed region is limited to the resonance region, where $m_{h_2}\simeq 2m_\chi$, as discussed in Sec.~\ref{sec:dm}. 
The region of $\Omega_\chi h^2\le 0.12$ becomes broadened as $\alpha$ gets larger. However, the maximum value of $v_S$ gets lowered, and the larger $m_{h_2}$ region is also more constrained by the perturbative unitarity, as seen in the lower-right plot. 
It should be noted that as long as $m_{h_2} = 2m_{\chi}$ is maintained, the allowed values of $m_{h_2}$ could vary, as shown in Fig.~\ref{fig:DM_vS_alp}. 
Taking all the constraints into account, we conclude that the parameter space that SKA could probe is limited only to the region, where $10~\text{TeV} \lesssim v_S\lesssim 200$ TeV with $m_{h_2}=2m_\chi$.

\begin{figure}[t]
\center
\includegraphics[width=9cm]{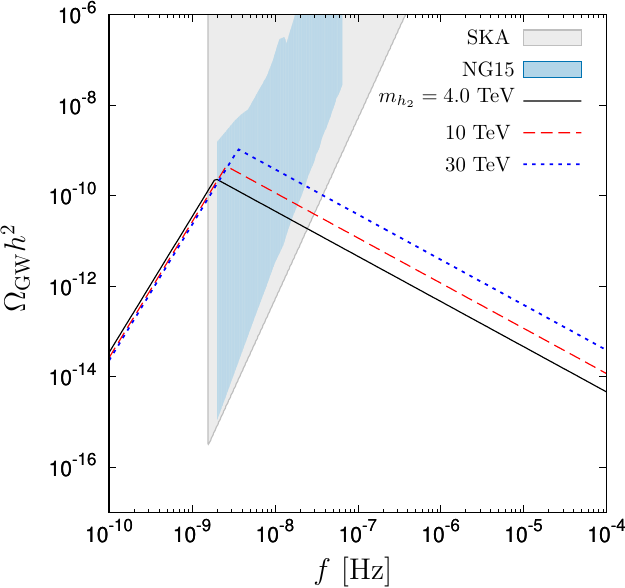}
\caption{$\Omega_{\text{GW}}h^2$ as a function of frequency $f$ in the cases of $m_{h_2}=4.0$ TeV (black, solid line), 10 TeV (red, dashed line), and 30 TeV (blue, dotted line), respectively. For all cases, we take $v_S = 100$ TeV and $\alpha=0.10^\circ$, and the DM mass is fixed to $m_\chi = m_{h_2}/2$ in order to satisfy $\Omega_\chi h^2\le \Omega_{\text{DM}}h^2=0.12$.
The grey-shaded region represents the SKA sensitivity, while the light-blue region (denoted as NG15) is indicated by the NANOGrav 15-year data.}
\label{fig:Oh2}
\end{figure}

\begin{table}[t]
\centering
  \begin{tabular}{|l|c|c|c|c|}  \hline
    & $\Omega_\text{GW}h^2(f_\text{peak})~[10^{-10}]$ & $f_\text{peak}~[10^{-9}~\text{Hz}]$ & $T_\text{ann}$ [MeV] & $\alpha_\text{DW}$ \\ \hline
    $m_{h_2}=4.0$ [TeV] & $2.4$ & $1.9$ & $17$ & $0.018$  \\ \hline
    $m_{h_2}=10$ [TeV] & $4.5$ & $2.6$  & $23$ & $0.025$ \\ \hline
    $m_{h_2}=30$ [TeV] & $10$ & $3.6$ & $32$ & $0.038$ \\ \hline
    \end{tabular}
    \caption{Summary of the GW-related parameters for the three benchmark points shown in Fig.~\ref{fig:Oh2}, where $\alpha_\text{DW}= \rho_\text{DW}(T_\text{ann})/\rho_r(T_\text{ann})$ with $\rho_\text{DW}$ and $\rho_r$ denoting the energy densities of DW and radiations, respectively.}
  \label{tab:gw}
\end{table}
In Fig.~\ref{fig:Oh2}, the values of $\Omega_{\text{GW}}h^2$ are displayed as a function of frequency $f$ for $m_{h_2}=4.0$ TeV (black, solid line), 10 TeV (red, dashed line), and 30 TeV (blue, dotted line). We use $v_S = 100$ TeV and $\alpha=0.10^\circ$, and maintain a fixed DM mass of $m_\chi = m_{h_2}/2$ to ensure that $\Omega_\chi h^2 \le \Omega_{\text{DM}}h^2 = 0.12$. The grey-shaded region represents the SKA sensitivity, while the light-blue shaded region is indicated by the NG15 data. 
As noted in Introduction, however, the ordinary DW interpretation is not favored since the best-fit low-frequency slope of GW spectrum reported by the NG15 data is $\Omega_\text{GW}\propto f^{1.2-2.4}$~\cite{NANOGrav:2023hvm,Babichev:2023pbf}, while $\Omega_\text{GW}\propto f^3$ in our case.\footnote{For {\it melting} DWs, $\Omega_\text{GW}\propto f^2$~\cite{Babichev:2023pbf}} Here, we consider the NG15 data as a constaint.  
The values of $\Omega_\text{GW}h^2(f_\text{peak})$, $f_\text{peak}$, $T_\text{ann}$, and $\alpha_\text{GW}$ in each case are summarized in Table~\ref{tab:gw}, where $\alpha_\text{DW}\equiv \rho_\text{DW}(T_\text{ann})/\rho_r(T_\text{ann})$ with $\rho_\text{DW}$ and $\rho_r$ representing the energy densities of DW and radiations, respectively.
Although $T_\text{ann}$ and $\alpha_\text{DW}$ in those cases lie outside the 95\% CL NG15-favored region, they are not ruled out~\cite{NANOGrav:2023hvm}.
Regardless of the small viable window for the DM relic abundance, GW arising from the DW collapses in the cxSM can accommodate $\Omega_\text{GW}h^2=\mathcal{O}(10^{-10}-10^{-9})$ in the nano-Hz frequency range, which can be further probed by the SKA experiment.

Before closing this section, we discuss a possible alteration in the DM relic density.
\begin{itemize}
\item If a significant amount of entropy is generated after the collapse of DW, the abundance of DM could be reduced~\cite{Hattori:2015xla}, which would expand the allowed region. However, in the parameter space we are examining, the energy density of DW is subdominant, and therefore, the large entropy production would not occur.
\item DM could be nonthermally produced after the collapse of DW. If DWs annihilate primarily into $h_2$, which subsequently decays into lighter particles including $\chi$, then the value of $\Omega_\chi$ would remain unchanged because $h_2\rightarrow \chi\chi$ is kinematically suppressed due to $m_\chi \simeq m_{h_2}/2$ in our allowed region. However, if $\chi$ is produced directly by the DW collapse, the value of $\Omega_\chi$ could be affected. To provide a quantitative assessment, we would need to have a complete understanding of the DW annihilation dynamics, which is outside the scope of our current study. We will defer this aspect to future work.
\end{itemize}

\section{Conclusion}\label{sec:conclusion}
We have conducted the study on GW signatures from DW collapses in the CP-conserving cxSM, in relation to DM physics. Our findings indicate that the bias term $|a_1|$, which is needed to make DW unstable, has to be greater than $\mathcal{O}(10^{-15})~\text{GeV}^3$ to be consistent with the BBN bound. 
Such a small value of $a_1$ results in $\sigma_\text{SI}^N(\propto a_1^2)$ being far below the latest LZ bound.
We also found that to have the region that can be probed by SKA, we should take $10~\text{TeV}\lesssim v_S \lesssim 200$ TeV and $1~\text{TeV}\lesssim m_{h_2}\lesssim 100$ TeV for a relatively small mixing angle $\alpha$, such as $\alpha=0.1^\circ$.
In such a parameter space, $\Omega_\chi$ tends to be overabundant due to the smallness of $\alpha$. Nevertheless, the allowed region can be marginally found if $m_\chi \simeq m_{h_2}/2$. If we take $\alpha$ to be larger, the region where $\Omega_\chi h^2<0.12$ gets broadened to some extent. However, the upper limit of $m_{h_2}$ becomes smaller owing to the perturbative unitarity constraint, diminishing the parameter space that gives detectable GW signatures. 
In any case, it is encouraging that future GW experiments may shed light on the higher-mass range of the additional scalar.

In this study, we focused on the minimal setup by dropping other U(1)-breaking terms. It would be interesting to see how much our results change quantitatively in more complicated cases. The analysis will be given elsewhere.


\appendix
\section{DM annihilation cross sections}\label{app:dm}
The DM annihilation processes (a)-(e) shown in Fig.~\ref{fig:DM} are, respectively, given by
\begin{align}
(\sigma v_{\text{rel}})_{\chi\chi\to h_{1,2}\to f\bar{f}}
& = N_C^f\frac{m_f^2\beta_{ff}^3}{16\pi v^2}
\left| \frac{\lambda_{h_1\chi\chi}\kappa_{1f}}{s-m_{h_1}^2+im_{h_1}\Gamma_{h_1}}
+\frac{\lambda_{h_2\chi\chi}\kappa_{2f}}{s-m_{h_2}^2+im_{h_2}\Gamma_{h_2}}
\right|^2, \label{sigmav_a}\\
(\sigma v_{\text{rel}})_{\chi\chi\to h_{1,2}\to VV}
& = \mathcal{S}\frac{m_V^4\beta_{VV}}{2\pi s v^2}\left[3+\frac{s^2\beta_{VV}^2}{4m_V^4}\right]
\left|
\frac{\lambda_{h_1\chi\chi}\kappa_{1V}}{s-m_{h_1}^2+im_{h_1}\Gamma_{h_1}}
+\frac{\lambda_{h_2\chi\chi}\kappa_{2V}}{s-m_{h_2}^2+im_{h_2}\Gamma_{h_2}}
\right|^2, \label{sigmav_b}\\
(\sigma v_{\text{rel}})_{\chi\chi\to h_{1,2}\to h_ih_j}
& = \mathcal{S}\frac{\beta_{h_ih_j}}{8\pi s}
\left|
\frac{\lambda_{h_1\chi\chi}\lambda_{h_1h_ih_j}}{s-m_{h_1}^2+im_{h_1}\Gamma_{h_1}}
+\frac{\lambda_{h_2\chi\chi}\lambda_{h_2h_ih_j}}{s-m_{h_2}^2+im_{h_2}\Gamma_{h_2}}
\right|^2, \label{sigmav_c}\\
(\sigma v_{\text{rel}})_{\chi\chi\to \chi \to h_ih_j}
& = \mathcal{S}\frac{\beta_{h_ih_j}}{\pi s}
	\frac{\lambda_{h_i\chi\chi}^2\lambda_{h_j\chi\chi}^2}{(s-m_{h_i}^2-m_{h_j}^2)^2}, \label{sigmav_d}\\
(\sigma v_{\text{rel}})_{\chi\chi\to h_ih_j}
&= \mathcal{S}\frac{\beta_{h_ih_j}}{8\pi s} \lambda_{h_ih_j\chi\chi}^2, \label{sigmav_e}
\end{align}
where $N_C^f = 3$ for quarks and $N_C^f = 1$ for leptons, and 
\begin{align}
s &= \frac{4m_\chi^2}{1-v_{\text{rel}}^2/4}, \quad 
\beta_{ij} = 
\sqrt{1-\frac{2(m^2_{i}+m^2_{j})}{s}+\frac{(m^2_{i}-m^2_{j})^2}{s^2}},
\end{align}
with $v_{\text{rel}}\simeq 0.3$. $\mathcal{S}=1/2$ if the final states are identical particles, otherwise $\mathcal{S}=1$. 
The Higgs couplings are, respectively, given by
\begin{align}
\lambda_{h_1\chi\chi} & = \frac{1}{2}(\delta_2vc_\alpha+d_2v_Ss_\alpha) = \frac{(m_{h_1}^2v_S+\sqrt{2}a_1)s_\alpha}{v_S^2}, \label{lam133} \\
\lambda_{h_2\chi\chi} &= \frac{1}{2}(-\delta_2vs_\alpha+d_2v_Sc_\alpha) = \frac{(m_{h_2}^2v_S+\sqrt{2}a_1)c_\alpha}{v_S^2}, \\
\lambda_{h_1h_1h_1} &= \frac{3}{2}
\Big[
	\lambda vc_\alpha^3
	+\delta_2s_\alpha c_\alpha(vs_\alpha+v_Sc_\alpha)
	+d_2v_Ss_\alpha^3
\Big], \\
\lambda_{h_1h_1h_2} &= \frac{1}{2}
\Big[
	-3\lambda vs_\alpha c_\alpha^2
	+\delta_2\big\{-v(s_\alpha^3-2s_\alpha c_\alpha^2)
	+v_S(c_\alpha^3-2s_\alpha^2c_\alpha)\big\}
	+3d_2v_Ss_\alpha^2c_\alpha
\Big],\\
\lambda_{h_1h_2h_2} &= \frac{1}{2}
\Big[
	3\lambda vs_\alpha^2 c_\alpha
	+\delta_2\big\{v(c_\alpha^3-2s_\alpha^2 c_\alpha)
	+v_S(s_\alpha^3-2s_\alpha c_\alpha^2)\big\}
	+3d_2v_Ss_\alpha c_\alpha^2
\Big], \\
\lambda_{h_2h_2h_2} &= \frac{3}{2}
\Big[
	-\lambda vs_\alpha^3
	+\delta_2s_\alpha c_\alpha(-vc_\alpha+v_Ss_\alpha)
	+d_2v_Sc_\alpha^3
\Big] \label{lam222}, \\
\lambda_{h_1h_1\chi\chi} & = \frac{1}{2}(\delta_2c_\alpha^2+d_2s_\alpha^2), \\
\lambda_{h_2h_2\chi\chi} &= \frac{1}{2}(\delta_2s_\alpha^2+d_2c_\alpha^2), \\
\lambda_{h_1h_2\chi\chi} & = \frac{1}{2}(-\delta_2+d_2)s_\alpha c_\alpha.
\end{align}
The total decay widths of $h_1$ and $h_2$ are 
\begin{align}
\Gamma_{h_1} 
&= c_\alpha^2\Gamma_{\text{tot}}^{\rm SM}+\Gamma_{h_1\to h_2h_2}+\Gamma_{h_1\to \chi\chi}
=\frac{c_\alpha^2\Gamma_{\text{tot}}^{\rm SM}}{1-\text{Br}_{h_1\to h_2h_2}-\text{Br}_{h_1\to \chi\chi}},\label{Gam1}\\
\Gamma_{h_2} 
&= s_\alpha^2\Gamma_{\text{tot}}^{\rm SM}+\Gamma_{h_2\to h_1h_1}+\Gamma_{h_2\to \chi\chi}
=\frac{s_\alpha^2\Gamma_{\text{tot}}^{\rm SM}}{1-\text{Br}_{h_2\to h_1h_1}-\text{Br}_{h_2\to \chi\chi}} \label{Gam2},
\end{align}
where $\Gamma_{\text{tot}}^{\rm SM}=4.1$ MeV~\cite{LHCHiggsCrossSectionWorkingGroup:2013rie} and
\begin{align}
\Gamma_{h_i\to \chi\chi} &= \frac{\lambda_{h_i\chi\chi}^2}{32\pi m_{h_i}}\sqrt{1-\frac{4m_\chi^2}{m_{h_i}^2}},
~\theta(m_{h_i}-2m_\chi),
\quad i=1,2,\\
\Gamma_{h_1\to h_2h_2} 
&= \frac{\lambda_{h_1h_2h_2}^2}{32\pi m_{h_1}}\sqrt{1-\frac{4m_{h_2}^2}{m_{h_1}^2}}~\theta(m_{h_1}-2m_{h_2}), \\
\Gamma_{h_2\to h_1h_1} 
&= \frac{\lambda_{h_1h_1h_2}^2}{32\pi m_{h_2}}\sqrt{1-\frac{4m_{h_1}^2}{m_{h_2}^2}}
~\theta(m_{h_2}-2m_{h_1}),
\end{align}
with $\theta(x)$ denoting the step function.

%
\bibliography{refs}
%

\end{document}